\documentclass[iop,apj]{emulateapj}

\usepackage{amssymb}
\usepackage{amsmath}
\usepackage{graphicx}
\usepackage{rotating}
\usepackage{natbib}

\newcommand\peryr{${\rm yr}^{-1}\ $}

\shorttitle{OGLE Atlas of Classical Novae}
\shortauthors{Mr\'oz et al.}

\begin{document}

\title{OGLE Atlas of Classical Novae \\II. Magellanic Clouds}

\author{P. Mr\'oz\altaffilmark{1}, A. Udalski\altaffilmark{1}, R. Poleski\altaffilmark{1,2}, I. Soszy\'nski\altaffilmark{1}, M.~K. Szyma\'nski\altaffilmark{1}, G. Pietrzy\'nski\altaffilmark{1}, \\ \L{}. Wyrzykowski\altaffilmark{1}, K. Ulaczyk\altaffilmark{1,3}, S. Koz\l{}owski\altaffilmark{1}, P. Pietrukowicz\altaffilmark{1}, and J. Skowron\altaffilmark{1}}
\email{pmroz@astrouw.edu.pl}
\altaffiltext{1}{Warsaw University Observatory, Al. Ujazdowskie 4, 00-478 Warszawa, Poland}
\altaffiltext{2}{Department of Astronomy, Ohio State University, 140 W. 18th Avenue, Columbus, OH 43210, USA}
\altaffiltext{3}{Department of Physics, University of Warwick, Coventry CV4 7AL, UK}

\begin{abstract}
The population of classical novae in the Magellanic Clouds was poorly known because of a lack of systematic studies. There were some suggestions that nova rates per unit mass in the Magellanic Clouds were higher than in any other galaxy. Here, we present an analysis of data collected over 16 years by the OGLE survey with the aim of characterizing the nova population in the Clouds. We found 20 eruptions of novae, half of which are new discoveries. We robustly measure nova rates of $2.4 \pm 0.8$~yr$^{-1}$ (LMC) and $0.9 \pm 0.4$~yr$^{-1}$ (SMC) and confirm that the $K$-band luminosity-specific nova rates in both Clouds are $2-3$ times higher than in other galaxies. This can be explained by the star formation history in the Magellanic Clouds, specifically the re-ignition of the star formation rate a few Gyr ago. We also present the discovery of the intriguing system OGLE-MBR133.25.1160 which mimics recurrent nova eruptions.
\end{abstract}

\keywords{novae, cataclysmic variables -- stars: statistics}

\section{Introduction}

The Magellanic Clouds are a pair of irregular galaxies close to the Milky Way. With precisely known distances and low foreground reddening, they serve as important laboratories in studies of variable stars. However, no systematic search and study of classical novae (CNe) in these two galaxies has been conducted. So far, 46 novae have been discovered in the Large Magellanic Cloud (LMC) and 17 in the Small Magellanic Cloud (SMC; Pietsch 2010)\footnote{See http://www.mpe.mpg.de/$\sim$m31novae/opt/index.php for a complete list.}, but our knowledge about many eruptions is only fragmentary. 

Despite the fact that extragalactic nova rates are much easier to measure than those in the Milky Way (e.g.,  Shafter 2008), the nova rates in the Magellanic Clouds are poorly constrained. Graham (1979) assessed an eruption frequency of ``between 2 and 3 novae per year'' in the LMC and ``1 nova every 3 or 4 years'' in the SMC. Capaccioli et al. (1990) estimated a nova rate of $2 \pm 1$ \peryr{} in the LMC. Della Valle (2002) quotes $2.5 \pm 0.5$ \peryr{} (LMC) and $0.7 \pm 0.2$ \peryr{} (SMC), suggesting that the nova rates per unit mass are higher than in other galaxies. The nova rates adopted by Capaccioli et al. and by Della Valle are based on Graham's early estimates, making it possible that the actual uncertainties are larger than what is reflected by the quoted errors. It appears that Della Valle (2002) arbitrarily doubled Graham (1979)'s nova rate for the SMC in order to make it consistent with later observations.

The question of whether or not nova rates depend on the stellar population, and hence the Hubble type of a parent galaxy, has been studied by many authors, often with contradictory results. Ciardullo et al. (1990) proposed using the galaxy's $K$-band luminosity as a proxy for the stellar mass and introduced $K$-band luminosity-specific nova rates (LSNRs). They found that LSNRs are independent of the galaxy's Hubble type. On the other hand, Della Valle et al. (1994) noted that late-type galaxies (like the LMC or M33) seem to be more prolific nova producers. Later studies (e.g., Shafter et al. 2000; Williams \& Shafter 2004; Coelho et al. 2008; G\"uth et al. 2010; Franck et al. 2012; Curtin et al. 2015 and references therein) did not confirm this result: LSNRs seem to be constant ($1-3$ per year per $10^{10}\ L_{\odot,K}$) for a range of galaxy masses and Hubble types, with the possible exception of the Magellanic Clouds, in which nova rates were poorly constrained. We refer interested readers to Shafter (2008) for a comprehensive review.

Here, we present the results of a systematic search for CNe in the Magellanic Clouds based on long-term photometric data from the Optical Gravitational Lensing Experiment (OGLE) survey. In Section \ref{sec:data}, we describe our data and sample selection. Individual objects from our sample are described in Sections \ref{sec:lmc} (LMC) and \ref{sec:smc} (SMC). Nova rates are calculated in Section \ref{sec:rates}. In Section \ref{sec:rne}, we present constraints on the numbers of recurrent novae (RNe) in the Clouds and report the discovery of an intriguing system, OGLE-MBR133.25.1160. Finally, in Section \ref{sec:disc}, we suggest that properties of the Magellanic Cloud nova population can be explained by the star formation history in these irregular galaxies.

\begin{table*}
\centering
\caption{List of the Large Magellanic Cloud novae recorded by the OGLE survey.}
\begin{tabular}{lr@{$^{\rm{h}}$}r@{$^{\rm{m}}$}r@{\fs}rr@{$^{\circ}$}r@{$'$}r@{\farcs}rlllcrrl}
\hline
\multicolumn{1}{c}{Designation} & \multicolumn{4}{c}{R.A.$_{\rm J2000.0}$} & \multicolumn{4}{c}{Decl.$_{\rm J2000.0}$} & Field & Star ID & \multicolumn{1}{c}{Eruption Date} & \multicolumn{1}{c}{Type} &\multicolumn{1}{c}{$t_2$ [d]}&\multicolumn{1}{c}{$t_3$ [d]}& \multicolumn{1}{c}{Other Names / Remarks}  \\ \hline
LMCN 1999-09a & 05 & 19 & 55 & 79 & -70 & 27 & 53 & 4 & LMC\_SC21 & 212370 & 1999 Sep 10 & D & 16 & 19 &  \\
LMCN 2001-08a* & 04 & 48 & 57 & 50 & -69 & 55 & 35 & 7 & LMC136.7 & 7N      & 2001 Aug 09 & C & -- & -- &  OGLE-2001-NOVA-01 \\
LMCN 2002-02a & 05 & 36 & 46 & 38 & -71 & 35 & 34 & 4 & LMC172.3 & 23898  & 2002 Feb 02 & -- & -- & -- &  \\
LMCN 2003-06a & 05 & 08 & 25 & 52 & -68 & 26 & 21 & 9 & LMC118.4 & 20317  & 2003 Jun 18 & -- & -- & -- &  \\
LMCN 2004-10a & 05 & 56 & 42 & 43 & -68 & 54 & 34 & 5 & LMC196.8 & 9115   & 2004 Oct 20 & S & 11 & 24 &  \\
LMCN 2005-09a & 06 & 06 & 36 & 55 & -69 & 49 & 34 & 6 & LMC206.3 & 44     & 2005 Sep 30 & P & 16 & 29 &  \\
LMCN 2005-11a & 05 & 10 & 32 & 68 & -69 & 12 & 35 & 7 & LMC111.6 & 9699   & 2005 Nov 22 & J & 79 & 112 &  \\
LMCN 2009-05a & 05 & 31 & 26 & 37 & -67 & 05 & 40 & 0 & LMC518.29 & 17226 & 2009 May 05 & -- & -- & -- &  \\
LMCN 2010-11a* & 05 & 09 & 58 & 40 & -71 & 39 & 52 & 7 & LMC508.01 & 1937  & 2010 Nov 21 & P & 3.5 & 5 & $P_{\rm orb} = 1.26432(8)$ d$\dagger$  \\
LMCN 2011-08a* & 05 & 43 & 48 & 46 & -69 & 19 & 31 & 0 & LMC553.13 & 35590 & 2011 Aug 14 & D & 9: & 14: &  OGLE-2011-NOVA-03 \\
LMCN 2012-03a & 04 & 54 & 56 & 82 & -70 & 26 & 56 & 3 & LMC530.19 & 3349  & 2012 Mar 26 & S & 3.4 & 5.0 & $P_{\rm orb} = 0.801460(3)$ d\\
LMCN 2012-10a** & 05 & 20 & 21 & 09 & -73 & 05 & 43 & 3 & LMC500.11 & 11N   & 2012 Oct 12 & S & 12 & 22 & OGLE-2012-NOVA-03 \\
LMCN 2013-10a** & 05 & 57 & 58 & 35 & -74 & 54 & 08 & 9 & LMC649.28 & 33N   & 2013 Oct 16 & S & 52 & 67 & OGLE-2013-NOVA-02 \\
\hline
\end{tabular}\\
\begin{flushleft}
* Novae discovered in the OGLE archival data.\\
** Novae discovered by the OGLE survey in real time.\\
$\dagger$ Data are taken from Mr\'oz et al. (2014).
\end{flushleft}
\label{tab:lmc}
\end{table*}

\section{Data}
\label{sec:data}

Photometric data analyzed in this paper were collected over the course of the three phases of the OGLE sky survey: OGLE-II (1997--2000; Udalski et al. 1997), OGLE-III (2001--2009; Udalski et al. 2008), and OGLE-IV (since 2010; Udalski et al. 2015). The majority of the observations were taken through the $I$-band filter with a typical cadence of 2--5 days. Currently, the survey  regularly monitors an area of 700 square degrees around the Magellanic Clouds and Bridge.

The sample of novae analyzed in this paper was selected in two steps. Firstly, we searched for brightenings in light curves of all of the objects in the OGLE photometric databases using the same procedure as was employed in Mr\'oz et al. (2015). We also checked objects from W. Pietsch \& F. Haberl's list$^{\arabic{footnote}}$, finding the additional outbursts LMCN~2009-05a, SMCN~2002-10a, and SMCN~2011-11a. Our sample consists of 20 CNe, of which half are OGLE-based discoveries.

In Tables \ref{tab:lmc} and \ref{tab:smc}, we provide details for each system: equatorial coordinates (for the epoch J2000.0), OGLE identification, eruption date, light curve type (according to the classification scheme by Strope et al. 2010), timescales $t_2$ and $t_3$, and orbital period (if measured).

The time-series photometry of objects presented in this paper are available to the astronomical community from the OGLE Internet Archive\footnote{ftp://ftp.astrouw.edu.pl/ogle/ogle4/NOVAE/MC}. The real-time photometry of future novae will be available on the webpage\footnote{http://ogle.astrouw.edu.pl/ogle4/transients} of the OGLE-IV Transient Detection System (Wyrzykowski et al. 2014).

\section{LMC novae}
\label{sec:lmc}

Thirteen CNe were observed in the LMC during the period 1997--2013 by the OGLE survey (see Tab. \ref{tab:lmc}). Five were discovered in the OGLE data (two were announced in real-time). The properties of the LMC novae were comprehensively summarized by Shafter (2013), and so only a few objects are described in this section.

\subsection{LMCN 2001-08a}

This star was serendipitously found during a search for gravitational microlensing events toward the Magellanic Clouds (Wyrzykowski et al. 2011). The light curve (Fig. \ref{fig:lcs}) resembles that of a binary microlens, but EROS photometric data (P. Tisserand, priv. comm.) show a sharp rise, followed by two maxima. The binary microlensing model does not fit the data (M. Penny, priv. comm.). This suggests that the transient was a CN of the rare C-type (C stands for cusp-shaped secondary maximum; see Strope et al. 2010 for definition). 

The eruption started between 2001 July 9 and August 9. After a month of smooth decline, the nova started to brighten slowly, reaching the second maximum at $I=13.3$~mag $\sim 50$ days later. This was followed by a sharp drop by 5~mag in 12~days, which is also typical for C-type novae. The progenitor of this nova is undetected in the OGLE template image, meaning $I>21.5$~mag.

\subsection{LMCN 2011-08a}

This nova was found in the OGLE archival data. The eruption took place between 2011 May 15 and August 14. The observed decline was fast with $t_2 \lesssim 9$ and $t_3 \lesssim 14$ d (however, the peak was not covered, so the true decline times might differ), which was followed by a dust minimum (D-type light curve; Fig. \ref{fig:lcs}). The progenitor of this nova is again undetected in the OGLE template image, meaning $I>21.5$~mag.

\begin{table*}[t]
\centering
\caption{List of the Small Magellanic Cloud novae recorded by the OGLE survey.}
\begin{tabular}{lr@{$^{\rm{h}}$}r@{$^{\rm{m}}$}r@{\fs}rr@{$^{\circ}$}r@{$'$}r@{\farcs}rlllcrrl}
\hline
\multicolumn{1}{c}{Designation} & \multicolumn{4}{c}{RA$_{\rm J2000.0}$} & \multicolumn{4}{c}{DEC$_{\rm J2000.0}$} & Field & Star ID & \multicolumn{1}{c}{Eruption Date} &\multicolumn{1}{c}{Type} &\multicolumn{1}{c}{$t_2$ [d]}&\multicolumn{1}{c}{$t_3$ [d]}&  \multicolumn{1}{c}{Other Names / Remarks}  \\ \hline
SMCN 2001-10a & 00 & 46 & 27 & 92 & -73 & 29 & 45 & 5 & SMC103.5 & 230N    & 2001 Oct 09 & S & 40 & -- & \\
SMCN 2004-06a* & 00 & 55 & 03 & 21 & -73 & 38 & 03 & 5 & SMC106.8 & 5899  & 2004 Jun 13 & J & -- & -- & OGLE-2004-NOVA-01 \\
SMCN 2005-08a & 01 & 15 & 00 & 07 & -73 & 25 & 37 & 6 & SMC116.2 & 222   & 2005 Aug 06 & S & 14 & 20 & \\
SMCN 2006-08a* & 00 & 57 & 03 & 87 & -73 & 10 & 11 & 6 & SMC106.5 & 20822 & 2006 Aug 03 & O & 31.5 & 59 & OGLE-2006-NOVA-02 \\
SMCN 2008-10a* & 00 & 55 & 37 & 74 & -72 & 03 & 14 & 4 & SMC108.6 & 37477 & 2008 Oct 31 & -- & 170 & -- & OGLE-2008-NOVA-02 \\
SMCN 2012-06a** & 00 & 32 & 55 & 06 & -74 & 20 & 19 & 7 & SMC714.19 & 5N   & 2012 Jun 05 & S & 40 & 75 & OGLE-2012-NOVA-02 \\
SMCN 2012-09a* & 00 & 48 & 07 & 28 & -72 & 46 & 34 & 0 & SMC719.13 & 8017 & 2012 Sep 17 & S & 7 & 12 & OGLE-2012-NOVA-04 \\ \hline
SMCN 2002-10a & 00 & 56 & 30 & 44 & -72 & 36 & 28 & 5 & SMC105.6 & 39517 & 2002 Oct 15 & -- & -- & -- & X-ray binary (?)\\
SMCN 2011-11a & 01 & 59 & 25 & 87 & -74 & 15 & 28 & 0 & MBR109.24 & 18   & 2011 Nov 11 & -- & -- & -- & Be star\\
\hline
\end{tabular}\\
\begin{flushleft}
* Novae discovered in the OGLE archival data.\\
** Novae discovered by the OGLE survey in real time.
\end{flushleft}
\label{tab:smc}
\end{table*}

\subsection{LMCN 2012-03a}

This nova was discovered by Seach et al. (2012). From the OGLE data, we can constrain the moment of the eruption: it must have taken place between 2012 March 25.12340 and 26.02861 UT. In the latter moment, the nova had $I=11.7$~mag, suggesting that it might have been caught during the rise to the maximum. Assuming that the maximum took place on March 26.0 UT, the decline times (in the $I$ band) are $t_2=3.4 \pm 0.2$ d and $t_2=5.0 \pm 0.2$~d (Fig. \ref{fig:lcs}).

LMCN 2012-03a was observed with many facilities over the whole range of electromagnetic spectrum. The results of such a campaign are summarized by Schwarz et al. (2015). The observational evidence (very fast decline, fast turn-on/off times of the super-soft phase, orbital period) suggests that this might be an RN with a massive white dwarf primary.

In quiescence, the star has $I=18.15$~mag and $V-I = 0.02$~mag. We confirm the presence of ellipsoidal modulations with an orbital period of 0.801460(3) d, in agreement with Schwarz et al. (2015). This star has been observed by the OGLE survey since 2001 and we have not detected any other eruptions, although it could have happened in a seasonal gap (lasting four months).

\section{SMC novae}
\label{sec:smc}

From 1997 to 2013, OGLE recorded eruptions of seven CNe in the SMC (Tab. \ref{tab:smc}). Among these eruptions, five are OGLE-based discoveries (one, OGLE-2012-NOVA-02, was announced in real time). We describe them in detail below. 

\subsection{SMCN 2001-10a}

This nova was discovered by William Liller in photographs taken on 2001 October 21.0857 and 21.0879 UT (Liller \& Pearce 2001). Della Valle et al. (2001) and Bosch et al. (2001) reported spectra showing strong, double-peaked emission lines of H~{\sc i}, Na~{\sc i}, Fe~{\sc ii}, and O~{\sc i}. The detailed spectroscopic evolution of the nova was described by Mason et al. (2001).

The OGLE photometry shows that the eruption started between 2001 October 8 and 9, i.e., much earlier than reported by Liller. For the next two weeks the star saturated our images until it faded back below $I\sim 12$~mag. We recorded a slow, broken power-law decline (Fig. \ref{fig:lcs}) of $t_2 \approx 40$ d.

\subsection{SMCN 2004-06a\protect\footnote{This object was first spotted in the OGLE data by T. Mazewski (priv. comm.) and was later independently revealed in our systematic search.} }

This nova was discovered in OGLE frames collected on 2004 June 13, when it reached $I\sim 16$~mag. The eruption must have taken place in the seasonal gap, between 2004 February 2 and June 13. We recorded six rebrightenings (secondary maxima) with amplitudes of up to 2~mag and lengths of a few days (Fig. \ref{fig:lcs}) superimposed on a smooth nova decline. The post-nova is bright ($I\approx19.1$~mag), but we have not detected any significant periodic variability in its light curve.

\subsection{SMCN 2005-08a}

This nova was discovered by William Liller on 2005 August 6.388 UT (Liller \& Monard 2005). Mason et al. (2005) obtained a spectrum of the transient on August 8.17 UT, confirming that it was a CN. The spectrum was dominated by Balmer, Fe~{\sc ii}, and Na~{\sc i} emission lines. Balmer lines were broad and asymmetric with FWHM of about 3200 km s$^{-1}$. The {\it Chandra} satellite observed this nova 219 days after the eruption, but no X-rays were detected (Schwarz et al. 2011). 

This star has been monitored by the OGLE survey since 2001 (Fig. \ref{fig:lcs}). The object was heavily overexposed in the image taken on August 7.35374 UT. No source was visible at the position of the nova in the earlier frame (August 2.38290 UT). We assessed $t_2 = 14 \pm 2$ d and $t_3 = 20 \pm 2$~d. In quiescence, the star is below the detection limit ($I>21.5$~mag).

\subsection{SMCN 2006-08a$^{\arabic{footnote}}$}

The first image during the eruption was taken on 2006 August 3.32 UT when the star was saturated. It was undetected in the previous frame, taken five days earlier. We recorded some low-amplitude (0.5~mag) oscillations in the light curve during the early decline (Fig. \ref{fig:lcs}). Decline times are $t_2 = 31.5 \pm 2.5$ and $t_3 = 59 \pm 1$~d. In the pre-eruption images, the nova is likely blended with a faint $\sim 21$~mag star, and so we were unable to characterize the pre-nova. 

\subsection{SMCN 2008-10a}

The eruption of this nova started between 2008 October 28 and 31. The rise to the peak was slow (Fig. \ref{fig:lcs}). It took 18 days to reach the maximum brightness of $I = 13.0$~mag. We noted a short pre-maximum halt between 2008 November 2 and 6. The decline was very slow with a timescale of $t_2 = 170 \pm 10$~d. We also recorded a high-amplitude ($\sim 1$~mag), semi-periodic (16--18 d) variability around the maximum. The photometric properties of this star resemble those of the slow nova OGLE-2015-NOVA-01 (Aydi et al., in prep.).

\begin{figure*}
\centering
\begin{tabular}{cc}
\includegraphics[width=0.38\textwidth]{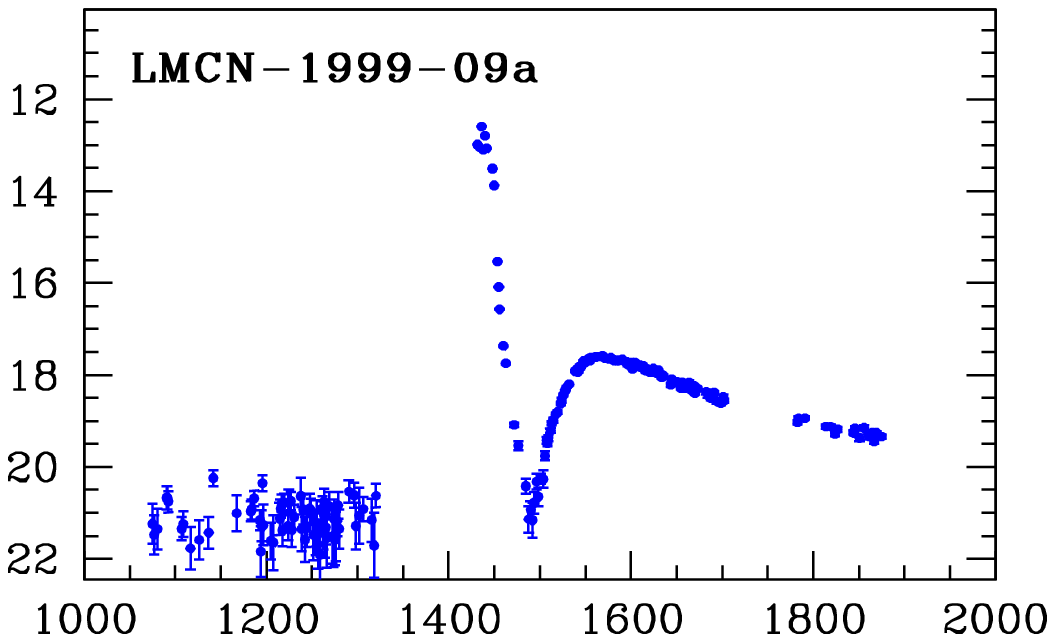} & 
\includegraphics[width=0.38\textwidth]{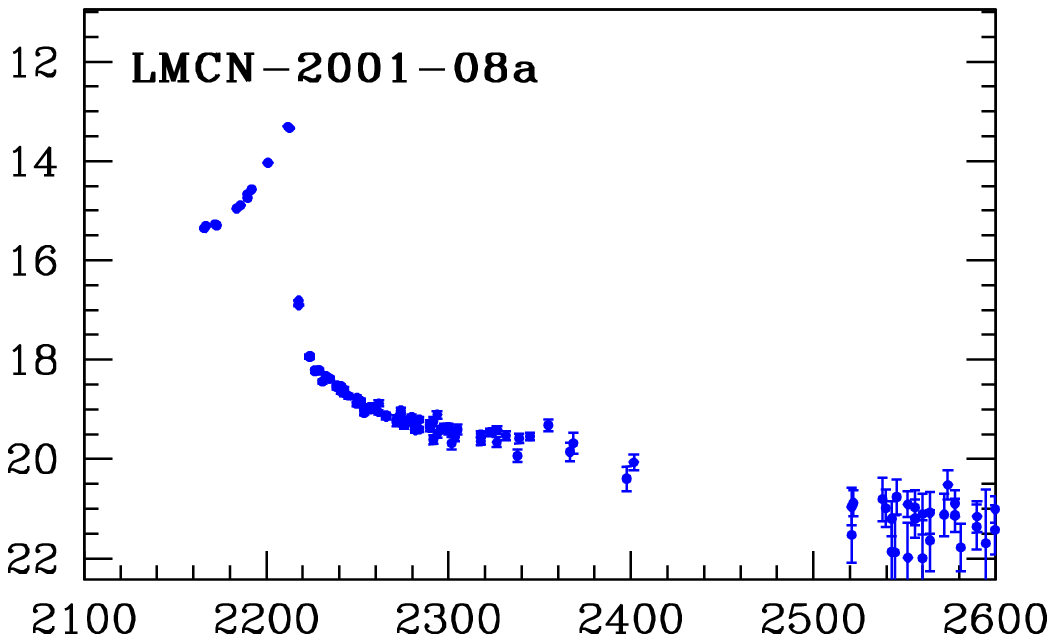} \\
\includegraphics[width=0.38\textwidth]{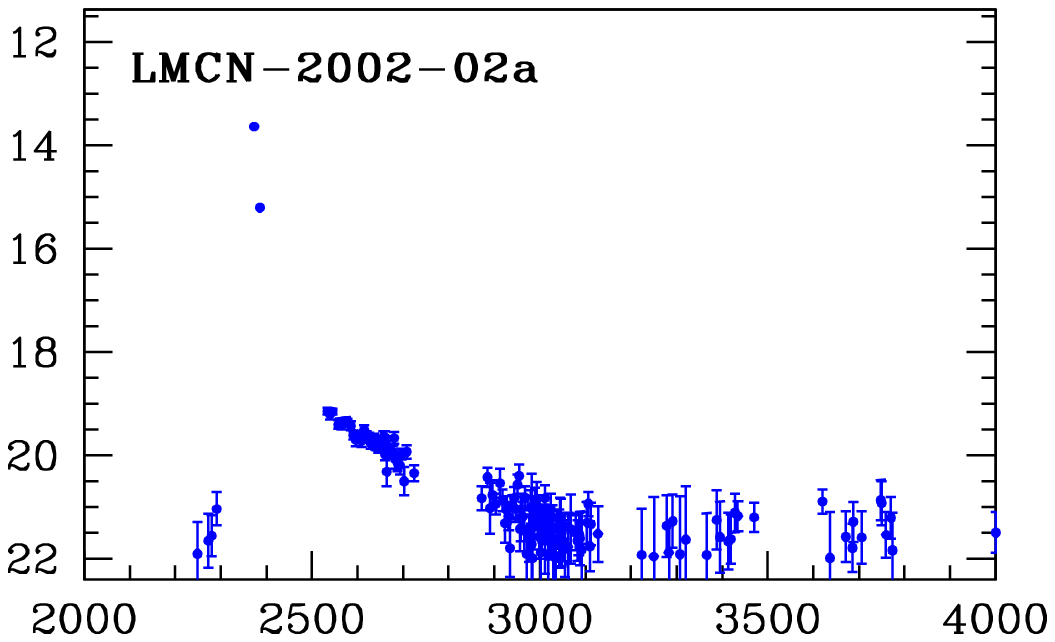} &
\includegraphics[width=0.38\textwidth]{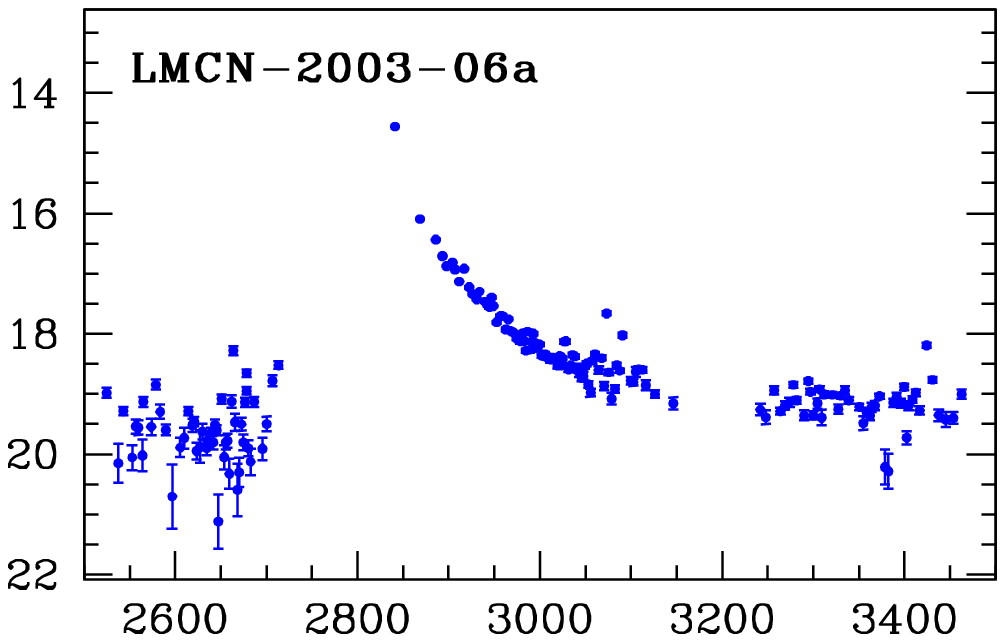} \\
\includegraphics[width=0.38\textwidth]{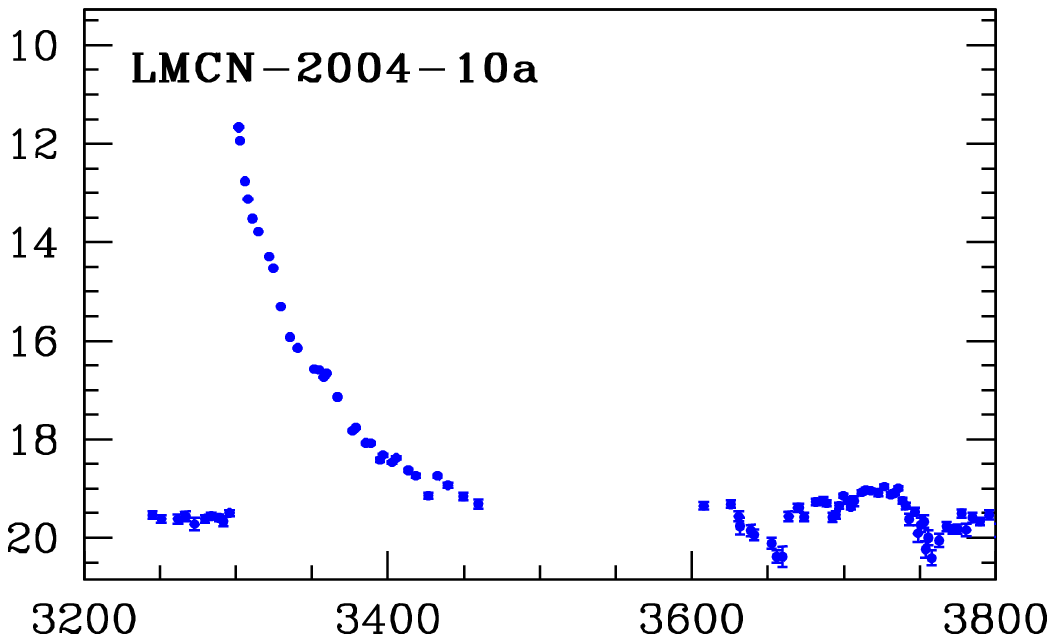} &
\includegraphics[width=0.38\textwidth]{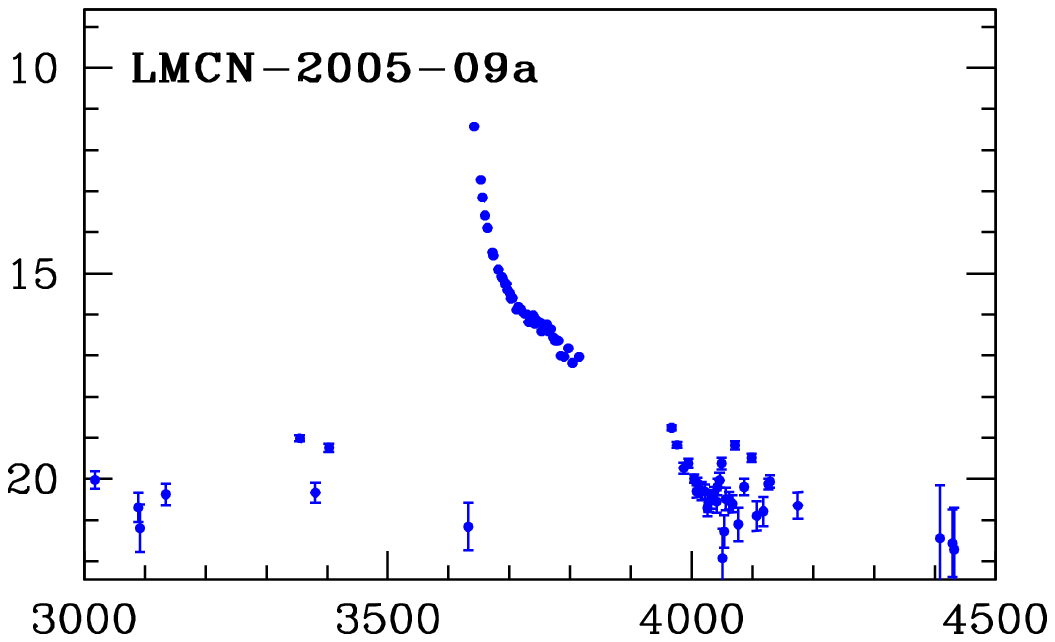} \\
\includegraphics[width=0.38\textwidth]{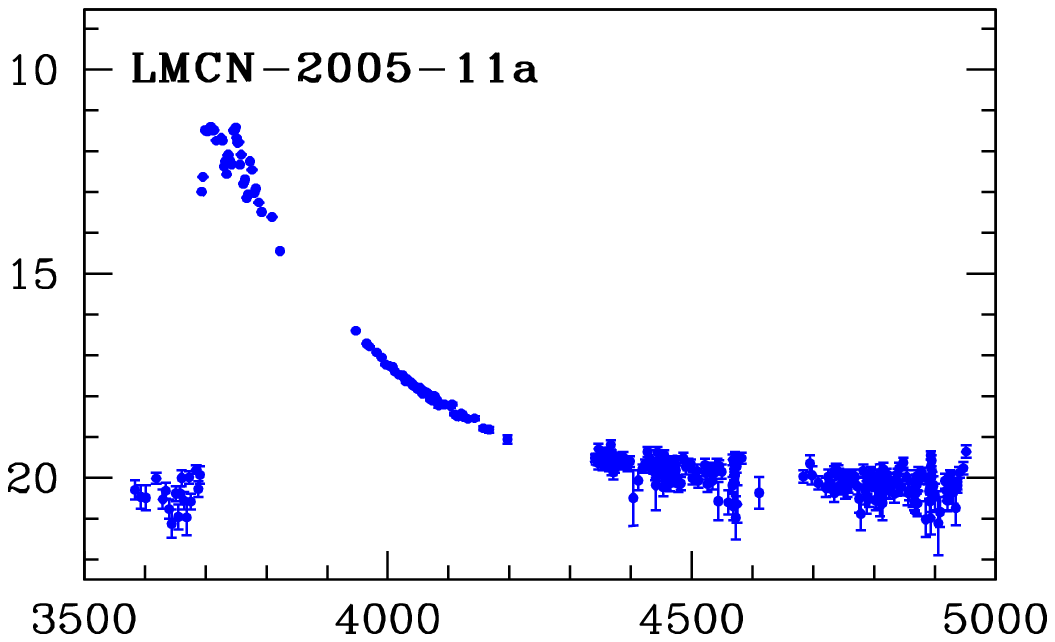} & 
\includegraphics[width=0.38\textwidth]{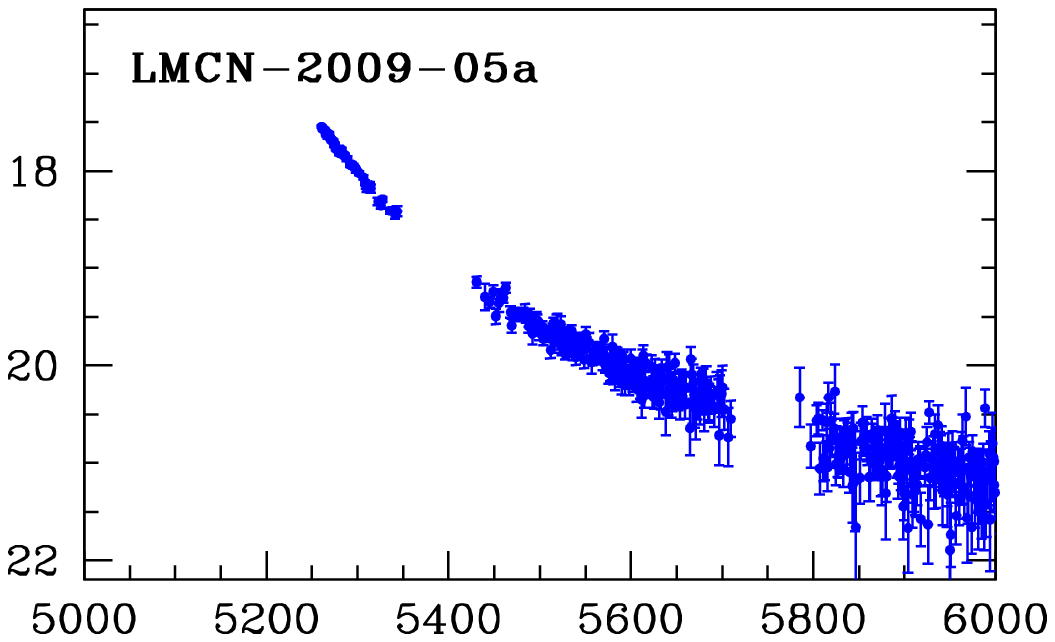} \\
\includegraphics[width=0.38\textwidth]{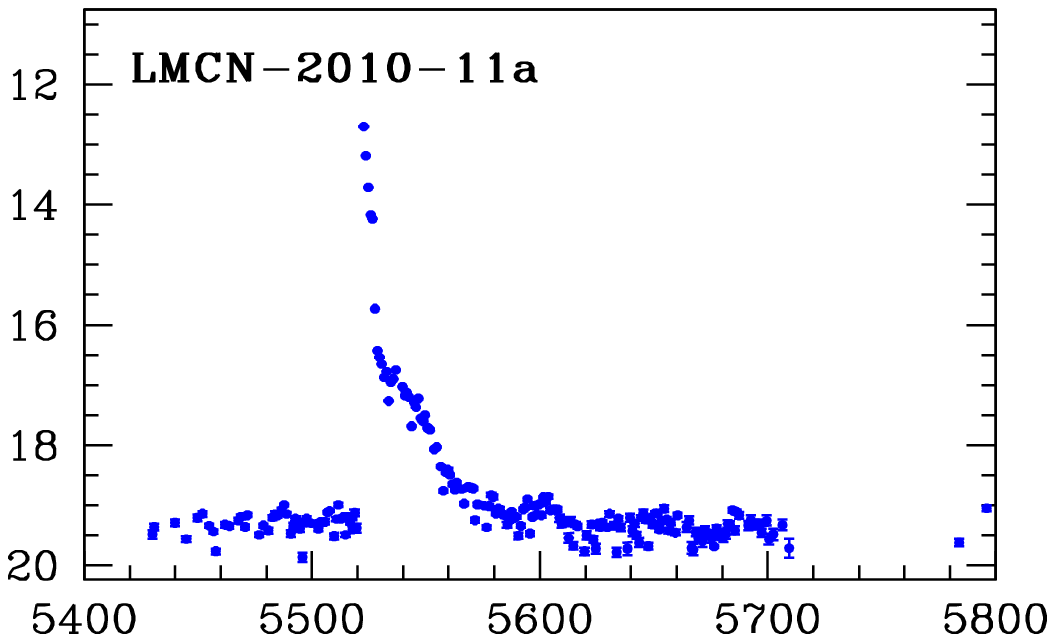} & 
\includegraphics[width=0.38\textwidth]{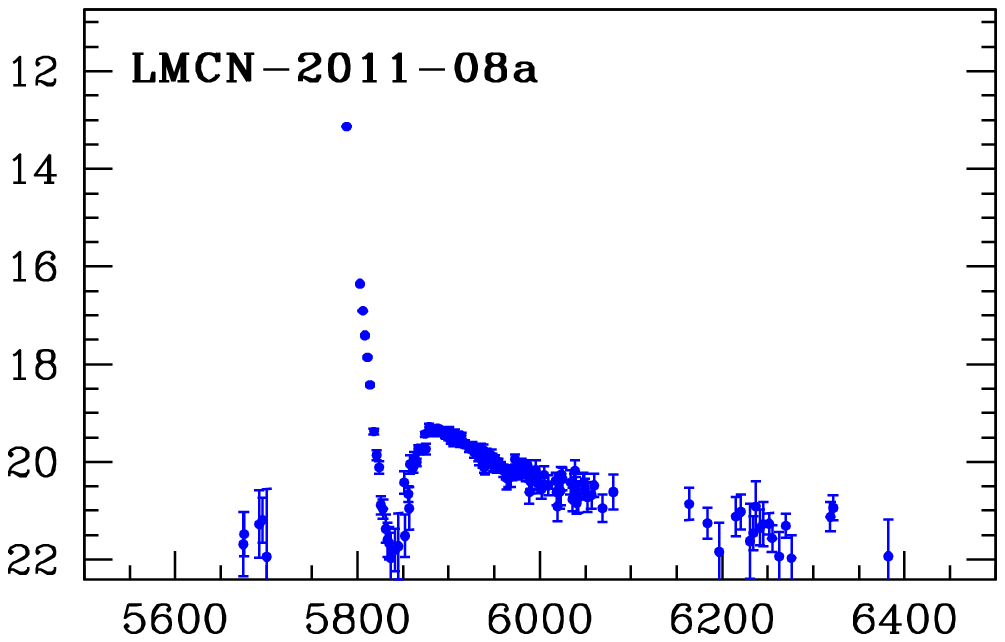} \\
\end{tabular}
\caption{Light curves of the Magellanic Clouds' novae. Time (in the Heliocentric Julian Date minus 2450000) is on the $x$-axis, the $I$-band magnitude is on the $y$-axis.}
\label{fig:lcs}
\end{figure*}
\addtocounter{figure}{-1}
\begin{figure*}
\centering
\begin{tabular}{cc}
\includegraphics[width=0.38\textwidth]{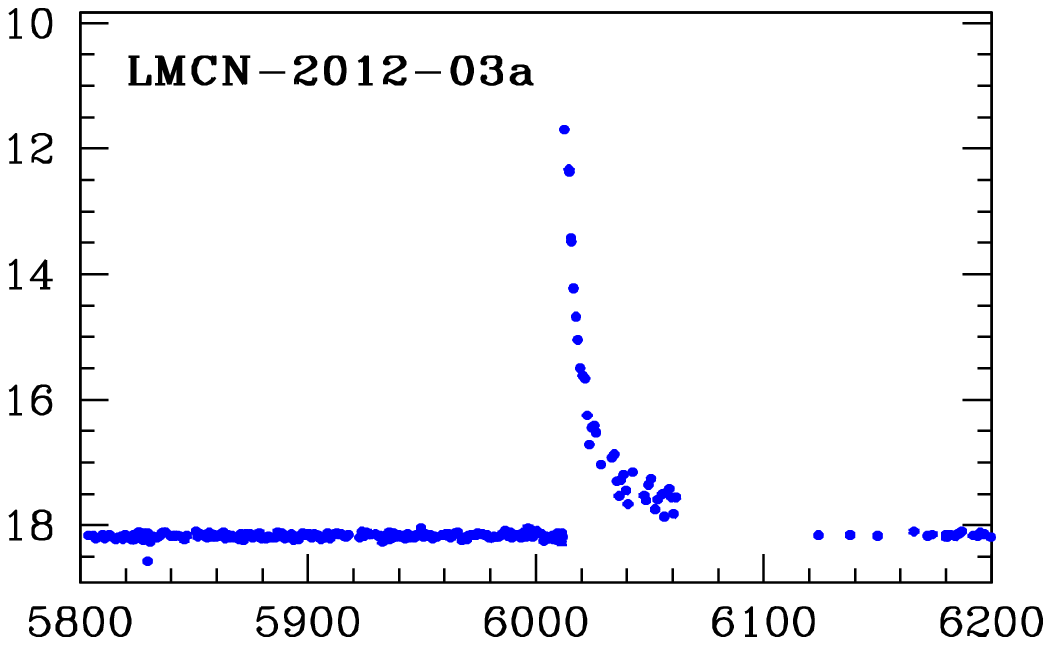} & 
\includegraphics[width=0.38\textwidth]{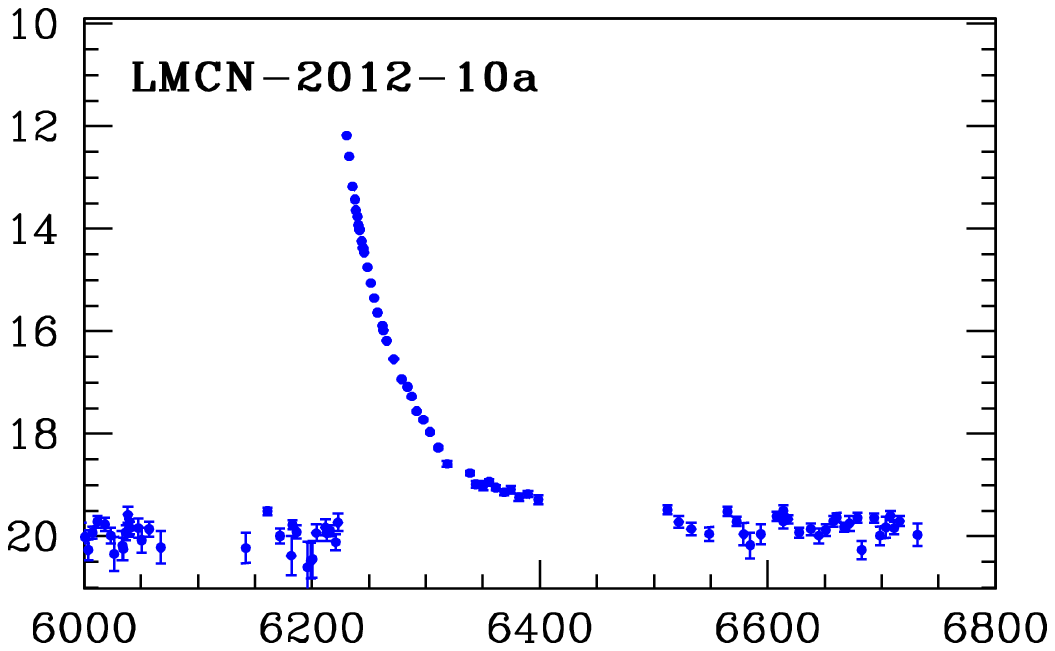} \\
\includegraphics[width=0.38\textwidth]{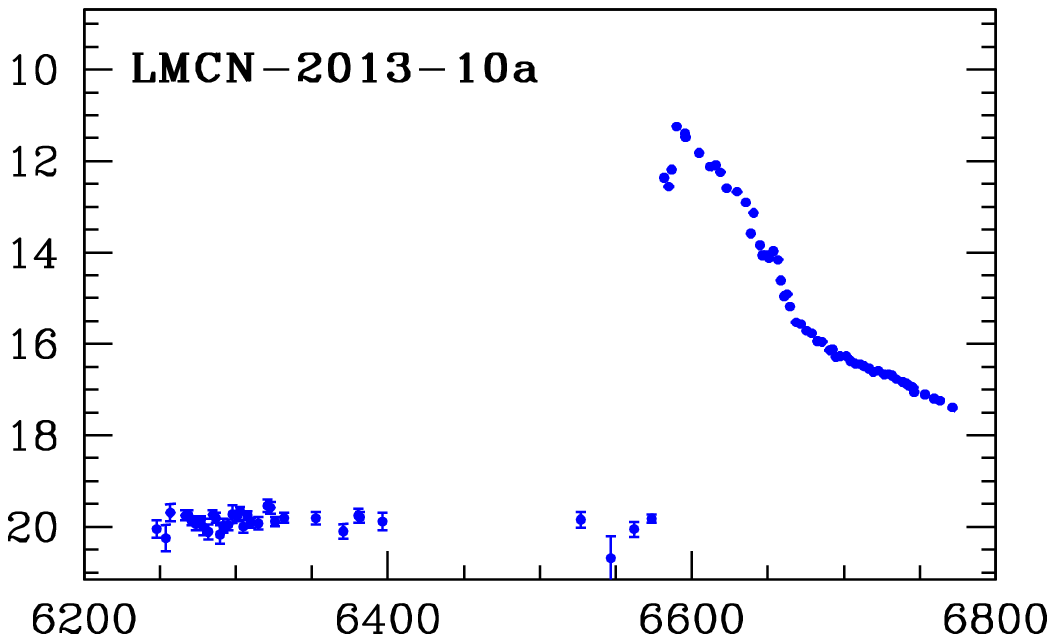} &
\includegraphics[width=0.38\textwidth]{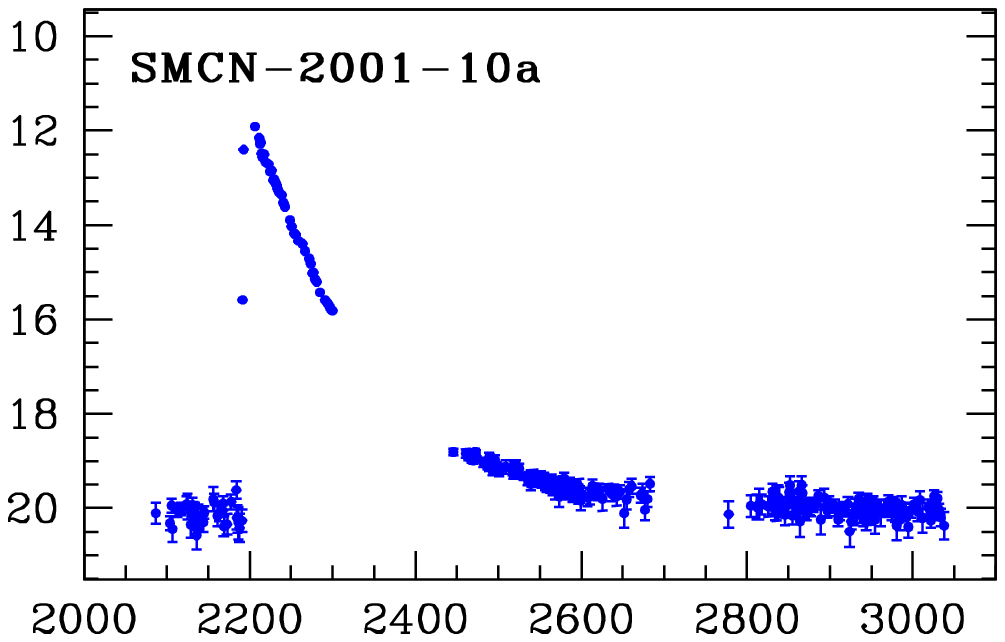} \\
\includegraphics[width=0.38\textwidth]{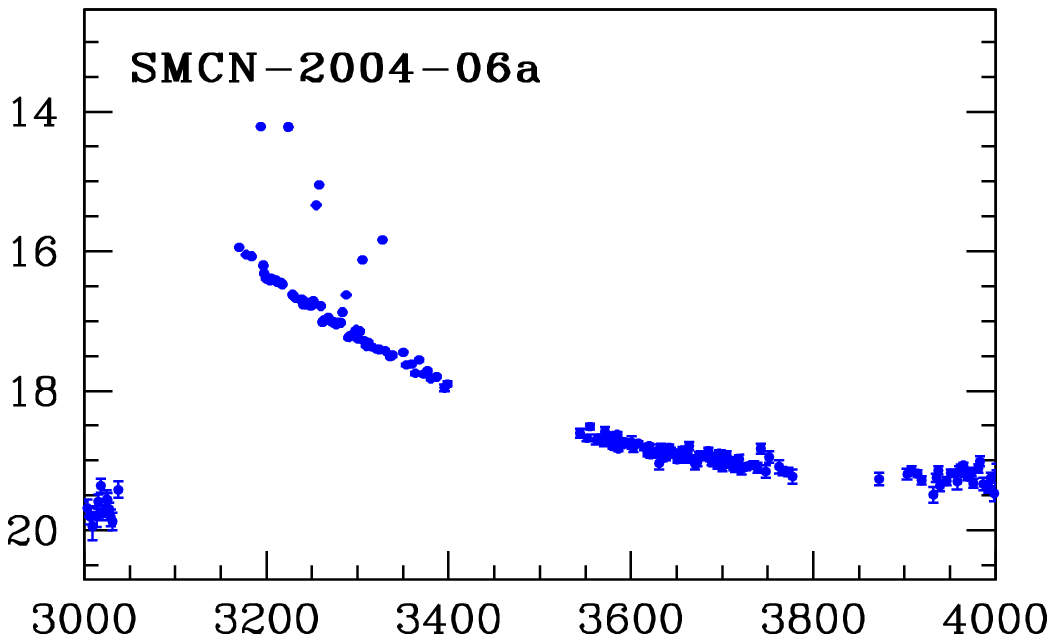} &
\includegraphics[width=0.38\textwidth]{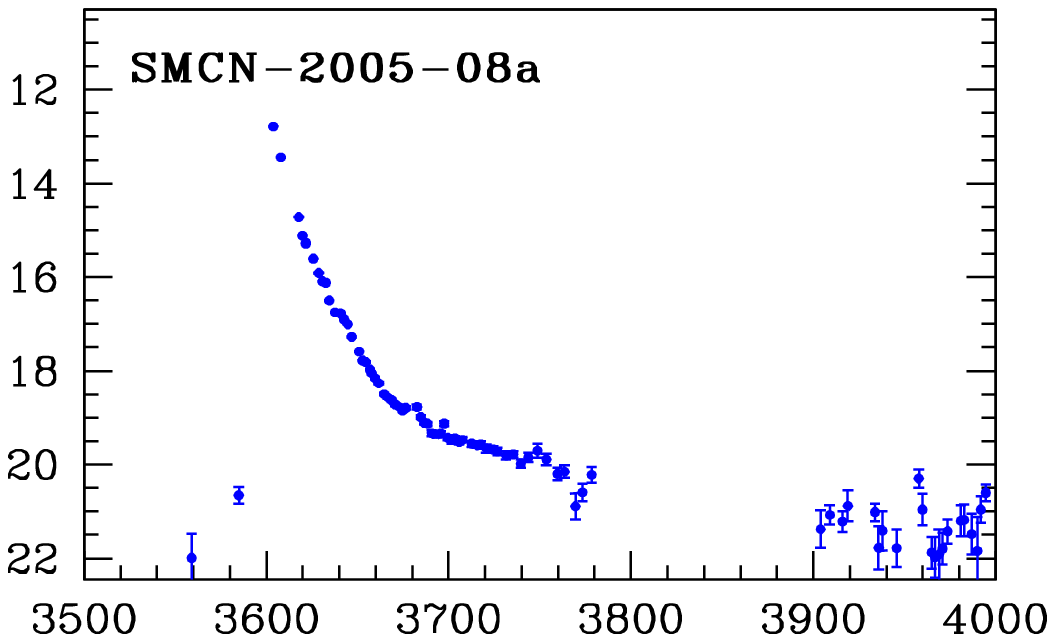} \\
\includegraphics[width=0.38\textwidth]{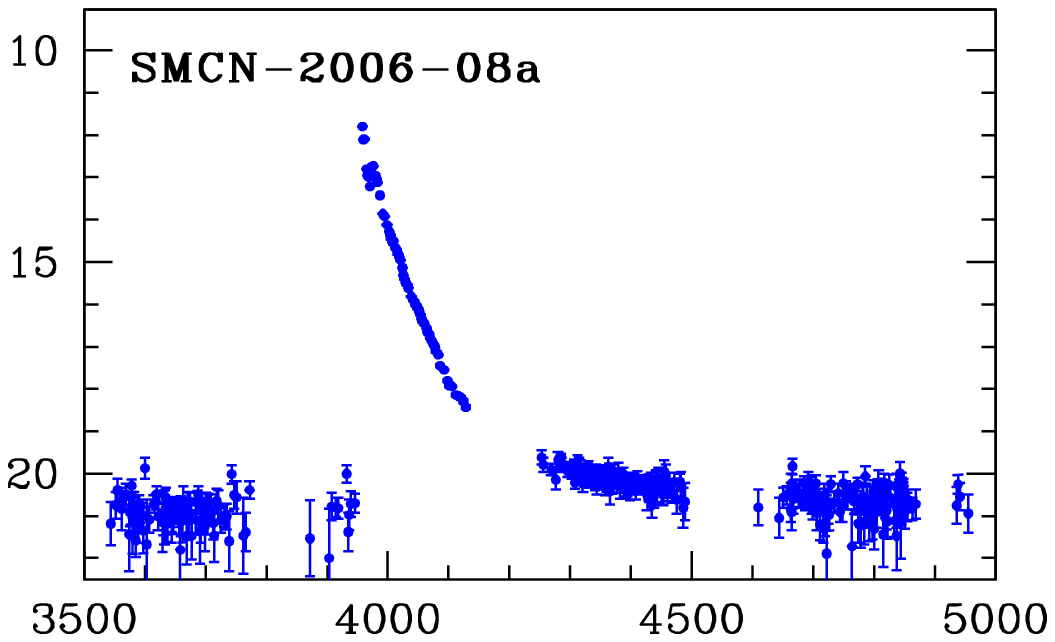} &
\includegraphics[width=0.38\textwidth]{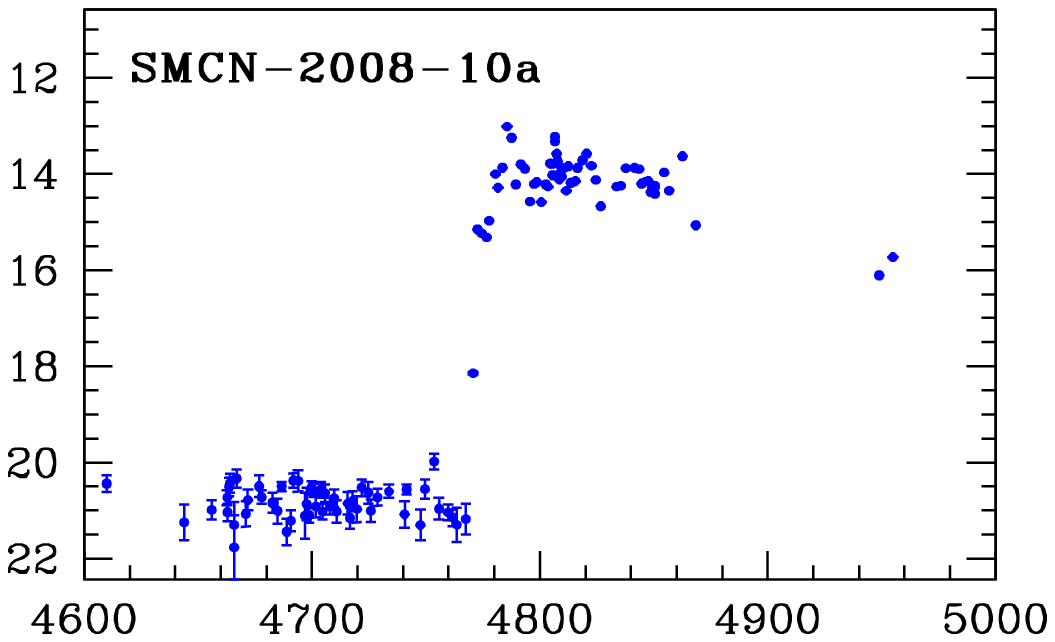} \\
\includegraphics[width=0.38\textwidth]{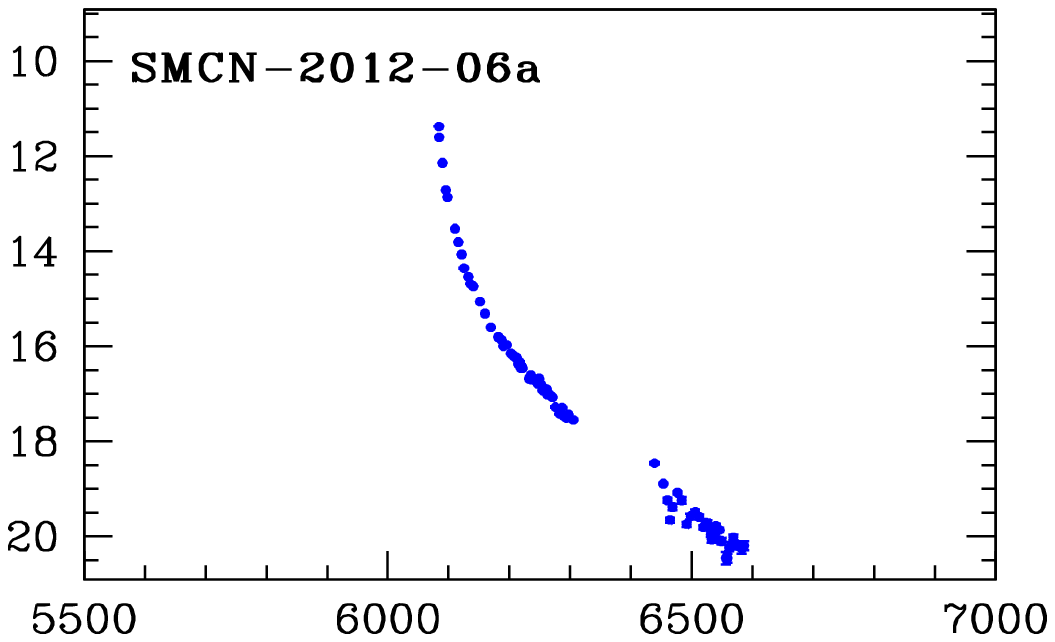} &
\includegraphics[width=0.38\textwidth]{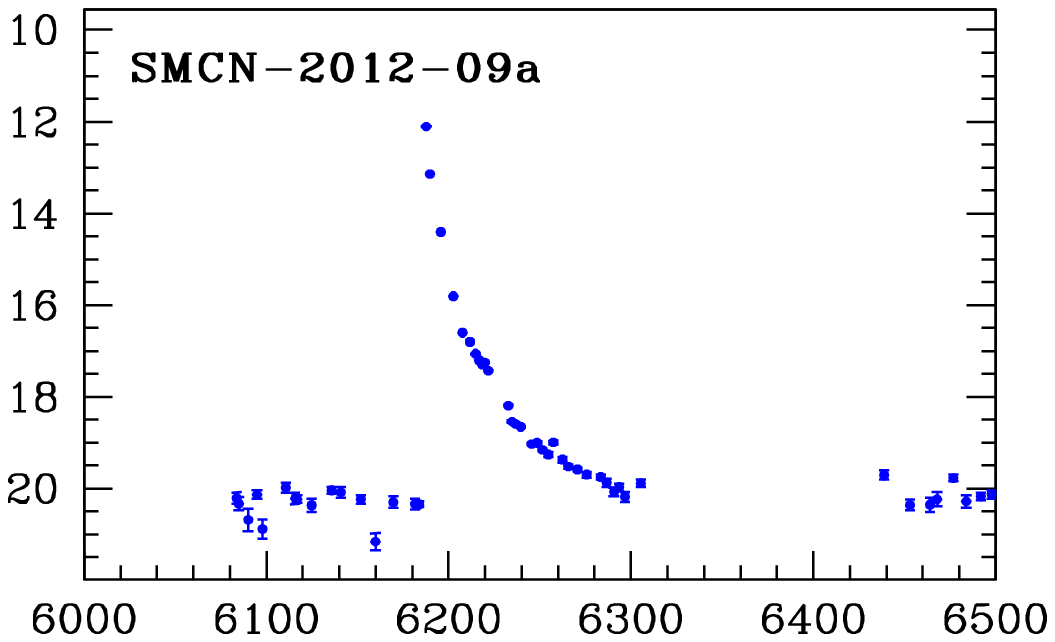} \\
\end{tabular}
\caption{(Continued.)}
\end{figure*}

\subsection{SMCN 2012-06a}

This nova was found in real time by the OGLE-IV Transient Detection System (Wyrzykowski et al. 2012, 2014). The eruption started between 2012 May 20 and June 5 (Levato et al. 2012). We assessed $t_2 \approx 40$ d and $t_3 \approx 75$ d (Fig. \ref{fig:lcs}). Schwarz et al. (2012) reported optical spectra confirming that the transient was a nova in its early nebular phase. They detected emission lines of H~{\sc i}, He~{\sc i}, [N~{\sc ii}], [Ne~{\sc iii}], and [O~{\sc iii}]. The H$\alpha$ line was asymmetric with a FWHM of 2400 km/s. Later optical spectra confirmed that the nova is of O--Ne type (Page et al. 2013a). The follow-up multiband photometry was carried out by Angeloni et al. (2012) and Unda-Sanzana et al. (2012). 

Follow-up {\it Swift} observations (2012 October) showed a very faint X-ray source with a count rate of $2.1^{+0.7}_{-0.6}\cdot 10^{-3}$ cts s$^{-1}$ in the 0.3--10 keV energy range (Masetti et al., Schwarz et al. 2012). Observations from 2013 March showed a much brighter X-ray source ($0.097 \pm 0.006$ cts s$^{-1}$) with a temperature of $64 \pm 3$ eV, indicating that the nova entered the super-soft (SS) phase (Page et al. 2013a). Page et al. (2013b) also reported the detection of a $0.850 \pm 0.013$ d period in the UV and X-ray data and interpreted it as the orbital period. 

\subsection{SMCN 2012-09a}

This nova was discovered in the archival OGLE-IV data. The first image during the eruption was taken on 2012 September 17.20~UT, when the star had $I\approx 12.1$~mag. The nova was very fast with $t_2 = 7 \pm 1$ d and $t_3 = 12 \pm 1$ d (Fig. \ref{fig:lcs}). This star has been monitored since 1997. We recorded some low-amplitude irregular variability in quiescence, but we have not found any periodic changes. The pre-nova is located redwards of the main sequence in the color-magnitude diagram, and so the secondary is likely a subgiant.

\subsection{SMCN 2002-10a and SMCN 2011-11a}

{\it SMCN 2002-10a}. The discovery of this transient was announced by Liller (2002). This star has been observed by the OGLE survey since 1997. The location of the star on the color-magnitude diagram ($I=15.70$~mag, $V-I=0.10$~mag) suggests that it might be a Be-type star (Mennickent et al. 2002). The quiescent light curve is dominated by long-term, irregular variations with an amplitude of 0.1~mag. There is additional short-period ($\sim 30$ d), low-amplitude ($\sim 0.05$~mag) sinusoidal variability. However, given the high amplitude of the outburst ($\sim3.5$~mag in $I$), it might have been a nova eruption in a binary with an early spectral type companion or an outburst in an X-ray binary. 

{\it SMCN 2011-11a}. This transient was detected by the X-ray telescopes MAXI (Kimura et al. 2011) and {\it Swift} (Kennea et al., Li et al., Li \& Kong 2011) as a soft X-ray source. The eruption of this star was likely caused by the nova explosion on the surface of a massive O-Ne-Mg white dwarf with a Be star companion (Morii et al. 2013; Ohtani et al. 2014).

\section{Nova rates}
\label{sec:rates}

In the years 2001--2013, we found 11 novae in the LMC and 7 in the SMC\footnote{We do not count LMCN 2009-05a (for which we only recorded a decline well after its 2009 eruption), SMCN 2002-10a, and SMCN 2011-11a (which are low-amplitude outbursts in Be stars).}, which gives observed rates of $0.9 \pm 0.3$ \peryr{} (LMC) and $0.5 \pm 0.2$ \peryr{} (SMC; we assumed a Poissonian error). Intrinsic nova rates were calculated similarly as for the Galactic bulge (Mr\'oz et al. 2015). We assumed that the nova distribution follows the light distribution (Shafter 2013 confirmed this for the LMC). Then, we created mock catalogs of novae drawn from photometric maps of the Magellanic Clouds, and finally calculated the probability of detection for each nova using the history of observations for a given field.

We began the analysis by drawing a random sample of stars brighter than $I_{\rm lim}=21$~mag from the OGLE-IV photometric database to create mock catalogs. We checked that the choice of $I_{\rm lim}$ does not influence the results. This is because of the high completeness of the OGLE photometry, which reaches 21.2~mag (Udalski et al. 2015). We assumed foreground Galactic contamination of $\sim 20,000$ stars per square degree, consistent with star counts in the northernmost OGLE fields LMC692--LMC694 and LMC701--LMC703, which are a good representation of the Galactic population (see Skowron et al. 2014).

We assessed the probability of detection of an eruption for each star from mock catalogs based on the temporal sampling of observations of the given fields. This allowed us to take into account observational bias: seasonal gaps, periods of bad weather, etc. We assumed that eruptions are distributed uniformly in time. Then, light curves from Fig. \ref{fig:lcs} were used as templates to create artificial data. The probability of detection determines the fraction of eruptions found by the same algorithm as that employed in the search procedure. It reached $\sim80$\% in the central regions of both galaxies. Overall, for the period 2001--2013, we were able to detect 35.5\% of novae in the LMC and 58.3\% in the SMC. This gives absolute nova rates of $2.4 \pm 0.8$ \peryr{} (LMC) and $0.9 \pm 0.4$ \peryr{} (SMC).

We checked the self-consistency of our measurements by using only the OGLE-IV observations (2010--2013). With five novae in the LMC and two in the SMC, the observed rates are $1.3 \pm 0.6$ \peryr{} and $0.5 \pm 0.4$ \peryr{}, respectively. Due to better spatial and temporal coverage, the detection probabilities are higher, 58.4\% (LMC) and 60.5\% (SMC), and so the true nova rates are equal to $2.1 \pm 1.0$ \peryr{} (LMC) and $0.8 \pm 0.6$ \peryr{} (SMC). These values are in good agreement with the results based on the longer timespan. Our quantitative analysis confirms estimates by Graham (1979) for the LMC.

\subsection{Luminosity-specific nova rates}
\label{sec:lsnr}

The nova rates in different galaxies can be compared with the aid of the $K$-band luminosity-specific nova rate $\nu_K$ (Ciardullo et al. 1990). For the majority of galaxies $\nu_K \approx 1-3$ per year per $10^{10}\ L_{\odot,K}$, although the uncertainties of the individual measurements can be large (e.g., Williams \& Shafter 2004). 

In the case of the Magellanic Clouds, the largest source of uncertainty is the $K$-band brightness of these galaxies. Williams \& Shafter (2004) recommend using values derived from integrated $B$-band brightness and $B-K$ color: $K_{\rm LMC} = -2.17 \pm 0.12$~mag and $K_{\rm SMC} = -0.43 \pm 0.15$~mag. Assuming a distance modulus to the LMC of $18.49 \pm 0.05$~mag (Pietrzy\'nski et al. 2013) and to the SMC of $18.95 \pm 0.07$~mag (Graczyk et al. 2014) and an absolute $K$-band brightness of the Sun of $+3.28$~mag (Binney \& Merrifield 1998), we obtain $L_K^{\rm {LMC}} = (3.7 \pm 0.5) \cdot 10^{9}\ L_{\odot,K}$ and $L_K^{\rm {SMC}} = (1.2 \pm 0.2) \cdot 10^{9}\ L_{\odot,K}$. However, the luminosities based on data from the 2MASS survey (Williams \& Shafter 2004) and the Diffuse Infrared Background Experiment (DIRBE) on board the {\it COBE} satellite (Israel et al. 2010) are systematically 25-50\% smaller, indicating that the $\nu_K$ found below are only lower limits.

For the Magellanic Clouds we find $\nu_K^{\rm {LMC}} = (6.5 \pm 2.2) \cdot 10^{-10}$ yr$^{-1}$ $L_{\odot,K}^{-1}$ and $\nu_K^{\rm {SMC}} = (7.5 \pm 3.1) \cdot 10^{-10}$ yr$^{-1}$ $L_{\odot,K}^{-1}$. These results corroborate previous suggestions (e.g., Della Valle et al. 1994; Williams \& Shafter 2004\footnote{We note that those authors rely on the Graham (1979) estimate of the nova rate of ``between 2 and 3 per year'' and their uncertainties of nova rates in the LMC and SMC are underestimated.}) that $\nu_K$ in low-mass Magellanic-like galaxies is 2--3 times larger than for high-mass spirals and ellipticals. 

\section{Recurrent novae}
\label{sec:rne}

There are two known RNe in the LMC (LMCN 1968-12a = LMCN 1990-02a = LMCN 2010-11a and LMCN 1937-11a = LMCN 2004-10a; Shafter 2013; Mr\'oz et al. 2014) and none in the SMC. Nova LMCN 2009-02a was suspected to be a second eruption of LMCN 1971-08a, but their positions do not match (see Shafter 2013; Mason \& Munari 2014). None of our objects show a second eruption (except the two mentioned earlier). We also checked the archival photometry from the MACHO survey (Alcock et al. 1993) spanning from 1993 to 1999, but we have not found any additional outbursts.

The OGLE survey is very sensitive to frequently erupting RNe. Following a procedure described in Mr\'oz et al. (2015), we estimated the probability of detecting of at least two eruptions for an RN with a recurrence timescale $\tau_{\rm rec}$ ranging from 300 d to 10 yr. We used a constant length of a detectability window $\Delta T$ of five or ten days. Our results are summarized in Table \ref{tab:rne}. For example, we should detect more than $\sim 2/3$ of all RNe with $\tau_{\rm rec} \lesssim 3$ yr. We have not detected any such novae (with the exception described below), practically ruling out their presence in the Magellanic System.

\begin{table}
\centering
\caption{Probability of detection of at least two eruptions of an RN with a recurrence timescale of $\tau_{\rm rec}$. $\Delta T$ is the length of a detectability window.}
\begin{tabular}{l|rr|rr}
$\tau_{\rm rec}$ & LMC & & SMC & \\
& $\Delta T = 10$ d & $\Delta T = 5$ d & $\Delta T = 10$ d & $\Delta T = 5$ d \\
\hline
300 d  & 76.1\% & 47.2\% & 86.8\% & 63.2\% \\
500 d  & 62.4\% & 22.9\% & 74.5\% & 43.6\% \\
1000 d & 59.4\% & 7.3\% & 75.0\% & 15.0\% \\
5 yr   & 16.0\% & 2.4\% & 27.9\% & 7.5\% \\
10 yr  &  8.0\% & 0.5\% & 16.5\% & 1.4\% \\
\hline
\end{tabular}
\label{tab:rne}
\end{table}

\begin{figure}
\includegraphics[width=.5\textwidth]{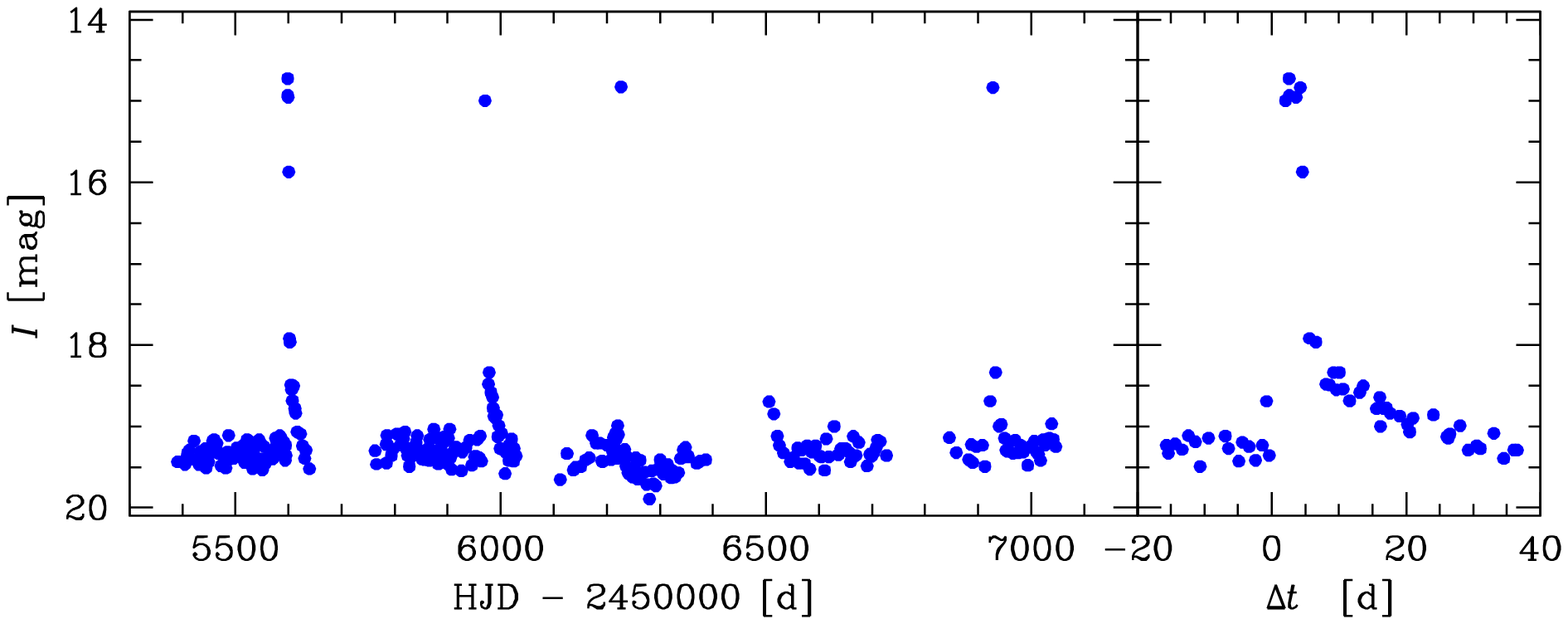}
\caption{Light curve of OGLE-MBR133.25.1160 showing fast, nova-like outbursts recurring on a timescale of $\sim330$ d.}
\label{fig1}
\end{figure}
\begin{figure}
\includegraphics[width=.5\textwidth]{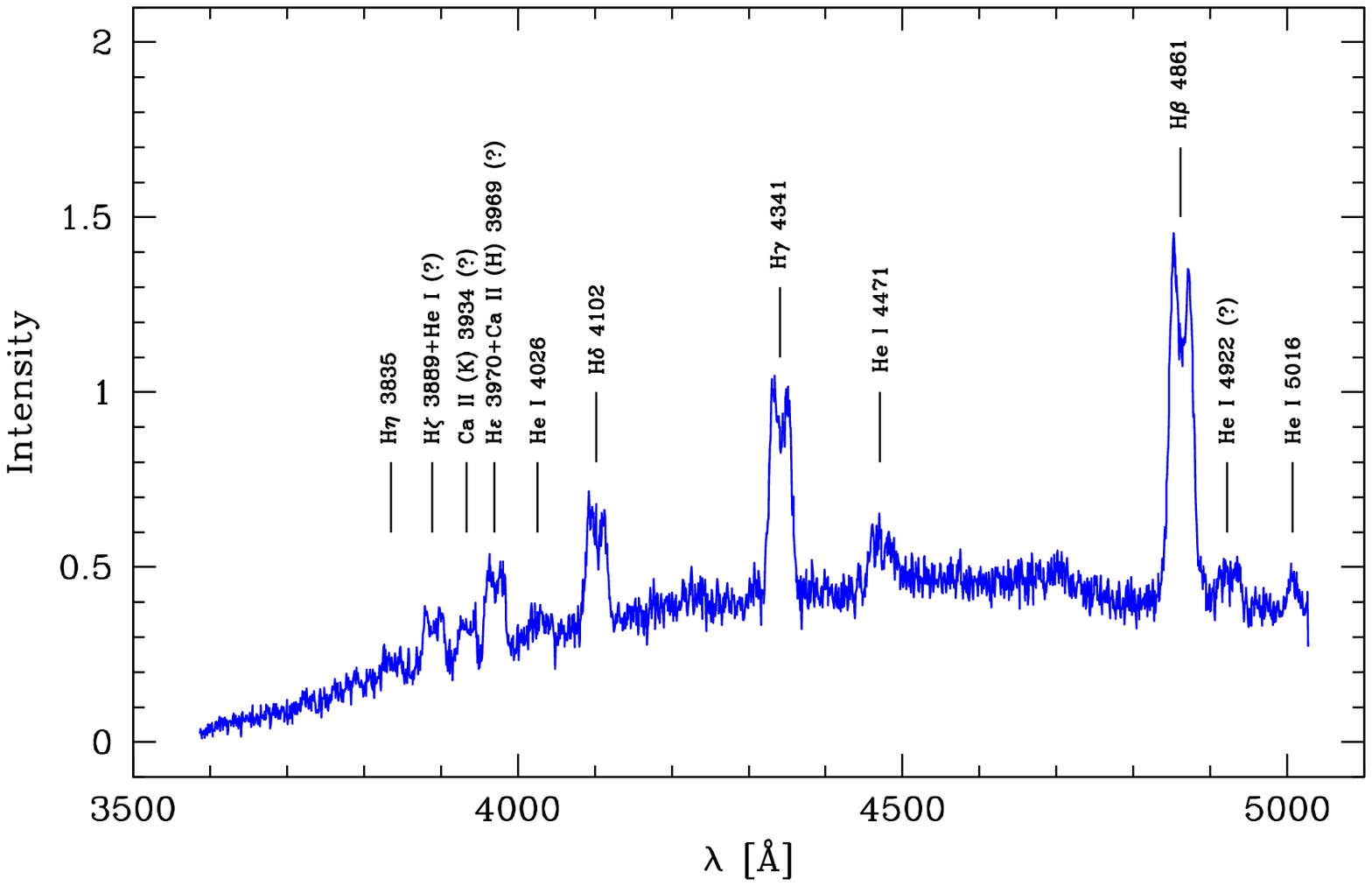}
\caption{Quiescent spectrum of OGLE-MBR133.25.1160 is a textbook example of a dwarf nova spectrum. It is dominated by the Balmer emission lines. It lacks strong He~{\sc ii} (4686 \AA) and other high excitation lines that should be present if the star were a RN.}
\label{fig}
\end{figure}

\subsection{The Curious Case of OGLE-MBR133.25.1160}
\label{sec:cc}

In our search for erupting variables toward the Magellanic Clouds, we found an interesting star: OGLE-MBR133.25.1160. We recorded its five outbursts during the period 2010--2014 with an average inter-eruption time of 330 d (see Fig. \ref{fig1}). Outbursts were very fast ($t_3 \approx 5$ d) and had an amplitude of at least 5~mag. The light curve does not resemble that of a typical dwarf nova. Dwarf nova outbursts have more symmetric light curves, often with a long ($\sim$several days) maximum. For example, Warner (2003) lists only two systems with inter-outburst times longer than 300~d, GK~Per and V1017 Sgr, and both of them have much slower outbursts. A rapid power-law decline observed in OGLE-MBR133.25.1160 bears some similarity to eruptions of CNe, which leads to the possibility that the star might be an RN with an ultra-short recurrence time (like M31 2008-12a; Darnley et al. 2014; Tang et al. 2014). 

The star is located toward the Magellanic Bridge ($\alpha=03^{\rm{h}}17^{\rm{m}}55{\fs}34$, $\delta=-73^{\circ}31'27{\farcs}3$). Its 2014 outburst was announced by the ASAS-SN survey as ASASSN-14ie (Shappee et al. 2014). Another outburst was recorded by the Catalina Sky Survey on 2006 September 18.7 UT (Drake et al. 2009).

To verify the nature of this star, spectroscopic observations were undertaken with the VLT/FORS2 spectrograph (Appenzeller et al. 1998) under program 295.D-5021 (PI: P. Mr\'oz). Spectra were taken in service mode on 2015 June 29/30 when the star was in quiescence. We obtained two spectra with exposure times of 1000~s each using the GRIS1200B+97 grism and a 1.3 arcsec slit width. The spectra cover the wavelength range 3600--5030 \AA~with a dispersion of 0.7 \AA~px$^{-1}$. They were reduced and wavelength calibrated using the ESO pipeline.

The final spectrum is shown in Fig. \ref{fig}. It is dominated by the hydrogen Balmer emission lines (H$\beta$ to H$\eta$) with some weaker lines of neutral helium and singly ionized calcium. Lines are double-peaked with velocities of $\pm 600-650$ km/s. This suggests that they originate from the accretion disk. 

There is no convincing evidence of He~{\sc ii} or other high excitation emission lines which are typical for post-novae and RNe in quiescence. In the spectra of RNe (e.g., U Sco or V394 CrA; Warner 2003), there are no Balmer lines because the hydrogen is likely fully ionized. 

We conclude that OGLE-MBR133.25.1160 is likely a Galactic dwarf nova mimicking RN eruptions. 

\section{Discussion and conclusions}
\label{sec:disc}

It is known that LMC novae tend to be faster than those from the Milky Way and M31 (e.g., Della Valle \& Duerbeck 1993, Shafter 2013), which is likely caused by relatively more massive white dwarfs in the nova progenitors. Indeed, we compared the distribution of $t_2$ times for the LMC and Milky Way bulge (Mr\'oz et al. 2015) novae using the Kolmogorov-Smirnov test and obtained a $p$-value of 0.03. 

We also compared the distribution of $t_2$ for SMC novae with those from the LMC and found a $p$-value of 0.36. However, neglecting the very slow nova SMCN 2008-10a gives larger $p=0.63$. After increasing the sample size with eight pre-2000 SMC novae\footnote{This should be treated with caution because decline times depend on the passband of the observations. Some measurements were taken in the $B$ band, others come from photographic plates.}, we obtained $p=0.54$. The average decline time is $\langle t_2\rangle = 22.4 \pm 8.6$ d for LMC novae and $\langle t_2\rangle = 26.5 \pm 6.9$ d for SMC novae ($22.6 \pm 3.7$ d after adding pre-2000 data). We thus conclude that the populations of SMC and LMC novae (in terms of decline times) are very similar. This indicates that the average white dwarf mass in nova binaries in the SMC is larger than for the M31 or Milky Way bulge.

Another common property of LMC and SMC novae is relatively high $K$-band LSNRs (Section \ref{sec:lsnr}), which are larger than in any other galaxy. Della Valle et al. (1994) suggested that high $\nu_K$ is a common feature of late-type, low-mass galaxies such as the Magellanic Clouds or M33. Yungelson et al. (1997) proposed that this is caused by the fact that the $K$-band luminosities of such galaxies do not reflect their mass, since they contain intermediate and young populations. The high value of $\nu_K$ for the Magellanic Clouds is also in line with the theoretical models of Matteucci et al. (2003), however, they fail to explain the low $\nu_K$ observed in elliptical galaxies. 
On the other hand, recent studies do not confirm the dependence of $\nu_K$ on the Hubble type of a parent galaxy (e.g. Williams \& Shafter 2004), except in the case of the Magellanic Clouds. 

According to Yungelson et al. (1997), the nova rate depends primarily on the star formation history. For a delta function star formation rate (SFR), the nova rate attains its peak after a few hundred Myr and then declines by a factor of 10 on a timescale of 5--10 Gyr. For a constant SFR, the nova rate reaches a constant level $\sim 3-4$ Gyr after the ignition of star formation (see Figs. 1 and 2 in Yungelson et al. 1997).

The star formation history is very similar for the LMC and SMC. After an initial star formation burst, the SFR was extinguished between about 12 and 5 Gyr ago, before reigniting at an approximately constant level $\sim 3-5$ Gyr ago (Harris \& Zaritsky 2004, 2009). Subramaniam \& Anupama (2002) analyzed the stellar populations in regions surrounding 15 novae in the LMC, finding that the majority of them belong to the intermediate-age population ($\sim 3 \pm 1$ Gyr). Stars younger than 3 Gyr constitute only $\sim 1/4$ of the total LMC mass (Harris \& Zaritsky 2009), and so the presence of intermediate and young populations cannot explain a $\nu_K$ higher than 2--3 times compared to the high-mass galaxies, as proposed Yungelson et al. (1997).

We thus suggest that properties of the Magellanic Clouds' novae, especially the high nova rates per unit mass, can be explained by the specific star formation history of both galaxies, i.e. the re-igniton of the SFR a few Gyr ago. 
The majority of novae were formed during the second, recent star formation event (5 Gyr ago or earlier). Being relatively young, white dwarfs in nova binaries should on average be more massive than in older systems. For example, LMC-metallicity stars with a ZAMS mass of $1.5 M_{\odot}$ or larger should end their evolution on the main sequence at a time of at most 2.4 Gyr. Moreover, because of the small metallicity of both Clouds (and weaker stellar winds during the star evolution), the mean mass of the white dwarfs should be slightly larger than for the same aged population with solar metallicity.  Novae hosting massive white dwarfs tend to be faster than those with lower-mass primaries, which explains the distribution of $t_2$ for both galaxies. Because high-mass white dwarfs need to accrete a smaller amount of gas to ignite a nova eruption (compared to low-mass ones), eruptions will be more frequent and thus nova rates will be enhanced.

If the SFR in the Magellanic Clouds extinguished 12 Gyr ago, their nova rates would be very small. For the LMC, after scaling the Yungelson et al. (1997) results, it should not exceed $0.5$ \peryr, i.e. $\nu_K \lesssim 2\cdot 10^{-10}$ yr$^{-1}$ $L_{\odot,K}^{-1}$, which would agree with the $\nu_K$ measured for the majority of galaxies. 
Re-ignition of the SFR a few Gyr ago caused an increase in the nova rates in both Magellanic Clouds to the observed values. According to Harris \& Zaritsky (2004, 2009), the typical SFRs during the last 3--5 Gyr were $\sim 0.1 M_{\odot}$ yr$^{-1}$ (SMC) and  $\sim 0.2 M_{\odot}$ yr$^{-1}$ (LMC). According to the Yungelson et al. (1997) models for a constant SFR, after a few Gyr, the nova rates should be roughly constant, and so the ratio of nova rates in the LMC and SMC should be roughly two, which agrees with the observed values.

Yungelson et al. (1997) used a few models with different initial mass ratio distributions $q$ for the binary components $f(q)\propto q^{\alpha}$. They suggest that the determination of the nova rates can be used to constrain the index $\alpha$. Indeed, our observations agree with the model's predictions for $-0.5<\alpha<0$.

We conclude that normalized nova rates are generally independent of parent galaxy type, with typical values in the range of $\nu_K \approx 1-3$ per year per $10^{10}\ L_{\odot,K}$. The Magellanic Clouds have $\nu_K$ 2--3 times higher. This discrepancy can be easily explained by the star formation history in these galaxies, namely, the re-ignition of the SFR a few Gyr ago.

\section*{Acknowledgments}

We thank the anonymous referee for providing useful comments that improved the clarity of this work. We acknowledge helpful discussions with Michael Shara.

P.M. is supported by the ``Diamond Grant'' No. DI2013/014743 funded by the Polish Ministry of Science and Higher Education.
The OGLE project has received funding from the National Science Center, Poland, grant MAESTRO 2014/14/A/ST9/00121 to A.U. 
This work has been supported by the Polish Ministry of Science and Higher Education through the program ``Ideas Plus'' award No. IdP2012 000162.
The VLT/FORS2 observations were performed at the European Southern Observatory (ESO), proposal 295.D-5021(A), as part of the Director's Discretionary Time.

\end{document}